\documentclass[aps,10pt,superscriptaddress]{revtex4}
\usepackage{graphicx}
\graphicspath{{./}{./IMAGES/}}
\usepackage{epstopdf}
\usepackage{amssymb}
\usepackage{mathrsfs}
\usepackage{amsmath}
\usepackage{color}
\usepackage{latexsym}
\usepackage{amsfonts}
\usepackage{hyperref}
\usepackage{subfigure}
\usepackage{wasysym}
\usepackage{dcolumn}
\usepackage{bm}
\usepackage[percent]{overpic}
\usepackage{natbib}
\usepackage{booktabs}
\usepackage{gensymb}

\begin{document}

\selectfont
\title{Numerical modelling of non-ionic microgels: an overview}

\author{Lorenzo Rovigatti}
\affiliation{Dipartimento di Fisica, {\em Sapienza} Universit\`a di Roma, Piazzale A. Moro 2, 00185 Roma, Italy}
\affiliation{CNR-ISC, Uos Sapienza, Piazzale A. Moro 2, 00185 Roma, Italy}

\author{Nicoletta Gnan}
\affiliation{CNR-ISC, Uos Sapienza, Piazzale A. Moro 2, 00185 Roma, Italy}
\affiliation{Dipartimento di Fisica, {\em Sapienza} Universit\`a di Roma, Piazzale A. Moro 2, 00185 Roma, Italy}

\author{Letizia Tavagnacco}
\affiliation{CNR-ISC, Uos Sapienza, Piazzale A. Moro 2, 00185 Roma, Italy}
\affiliation{Dipartimento di Fisica, {\em Sapienza} Universit\`a di Roma, Piazzale A. Moro 2, 00185 Roma, Italy}

\author{Angel J. Moreno}
\affiliation{Centro de F\'isica de Materiales (CSIC, UPV/EHU) and Materials Physics Center MPC, Paseo Manuel de Lardizabal 5, 20018 San Sebasti\'an, Spain}
\affiliation{Donostia International Physics Center, Paseo Manuel de Lardizabal 4, 20018 San Sebastian, Spain}

\author{Emanuela Zaccarelli}
\affiliation{CNR-ISC, Uos Sapienza, Piazzale A. Moro 2, 00185 Roma, Italy}
\affiliation{Dipartimento di Fisica, {\em Sapienza} Universit\`a di Roma, Piazzale A. Moro 2, 00185 Roma, Italy}

\definecolor{corr14okt}{rgb}{0,0,1}
\definecolor{moved}{rgb}{0,0,0}

\begin{abstract}
Microgels are complex macromolecules. These colloid-sized polymer networks possess internal degrees of freedom and, depending on the polymer(s) they are made of, can acquire a responsiveness to variations of the environment (temperature, pH, salt concentration, \textit{etc.}). Besides being valuable for many practical applications, microgels are also extremely important to tackle fundamental physics problems. As a result, these last years have seen a rapid development of protocols for the synthesis of microgels, and more and more research has been devoted to the investigation of their bulk properties. However, from a numerical standpoint the picture is more fragmented, as the inherently multi-scale nature of microgels, whose bulk behaviour crucially depends on the microscopic details, cannot be handled at a single level of coarse-graining. Here we present an overview of the methods and models that have been proposed to describe non-ionic microgels at different length-scales, from the atomistic to the single-particle level. We especially focus on monomer-resolved models, as these have the right level of details to capture the most important properties of microgels, responsiveness and softness. We suggest that these microscopic descriptions, if realistic enough, can be employed as starting points to develop the more coarse-grained representations required to investigate the behaviour of bulk suspensions.
\end{abstract} 

\maketitle

\section{Introduction}

Microgels are colloid-sized polymer networks that are important not only for industrial and biomedical applications, but also as model systems to investigate fundamental problems in condensed matter physics~\cite{fernandez2011microgel}. Nowadays there exist many established synthesis techniques that make it possible to generate microgels with different sizes, shapes and microscopic architecture~\cite{pelton2000temperature,fernandez2011microgel,geisel2015hollow,crassous2015anisotropic,schmid2016multi}. Of course, the properties of the final object do not depend only on the synthesis protocol, but also on the nature of the polymeric constituents~\cite{saunders1999microgel,fernandez2011microgel}. The possibility of changing the latter is of extreme importance, as using different polymers alters not only the topology of the final network, but also the way the single microgels interact with the environment and with themselves. 

The library  of possible systems has considerably grown in the past years, and it presently encompasses microgels that can respond to changes of, \textit{e.g.}, temperature~\cite{pelton2000temperature,stieger2004small}, pH~\cite{saunders1999microgel,nigro2015dynamic}, salt concentration~\cite{lopez2007macroscopically}, external fields~\cite{colla_likos}. The specific way with which the polymer network adjusts itself to a variation of the external conditions depends on the properties of the particles, but the most prominent effect is nearly invariably an overall change of particle size. The resulting \textit{swelling/deswelling} transition~\cite{fernandez2011microgel} is the main reason why microgels have become important for both applications and fundamental science. However, from the theoretical point of view, modelling such a transition at the microscopic scale is a formidable challenge, as the fine details that underlie the swelling or deswelling of a particular microgel depend not only on the environment but also on the specific experimental protocol employed for the synthesis. Indeed, the polymer-colloid duality~\cite{lyon2012polymer} of microgel particles grants them an inherent multi-scale nature that is hard to tackle. As a result, huge advances in the experimental synthesis of microgels have not been accompanied by comparable progresses in the development of theoretical and numerical models. In fact, it is only recently that detailed and realistic models have appeared, mainly thanks to the deeper knowledge of the inner structure of microgel systems acquired through careful experiments and to the rapid increase of available computational power and numerical tools. 

Surprisingly, despite the growing interest, a comprehensive summary of recent progresses on the numerical modelling of microgels is not presently available. Here we fill this gap by reviewing the simulation work done to characterise non-ionic (uncharged) microgels and microgel suspensions at different time- and length-scales, going from atomistic models to very coarse-grained pair-potential descriptions. If not explicitly stated otherwise, we will use microgels made of poly(N-isopropylacrylamide) (PNIPAM)~\cite{pelton1986preparation}, which is the most common polymer used in the synthesis of non-ionic microgels, as the experimental reference system. The overview discusses models and results obtained with atomistic (Section~\ref{sec:atomistic}),  monomer-resolved (Section~\ref{sec:monomers}) and more coarse-grained (Section~\ref{sec:coarse-graining}) descriptions. In Section~\ref{sec:perspectives} we discuss possible future developments that we deem as promising, while the last section contains our conclusive remarks.

\section{Atomistic simulations}
\label{sec:atomistic}

\begin{figure*}[t!]
\centering
\includegraphics[width=0.7\textwidth,clip]{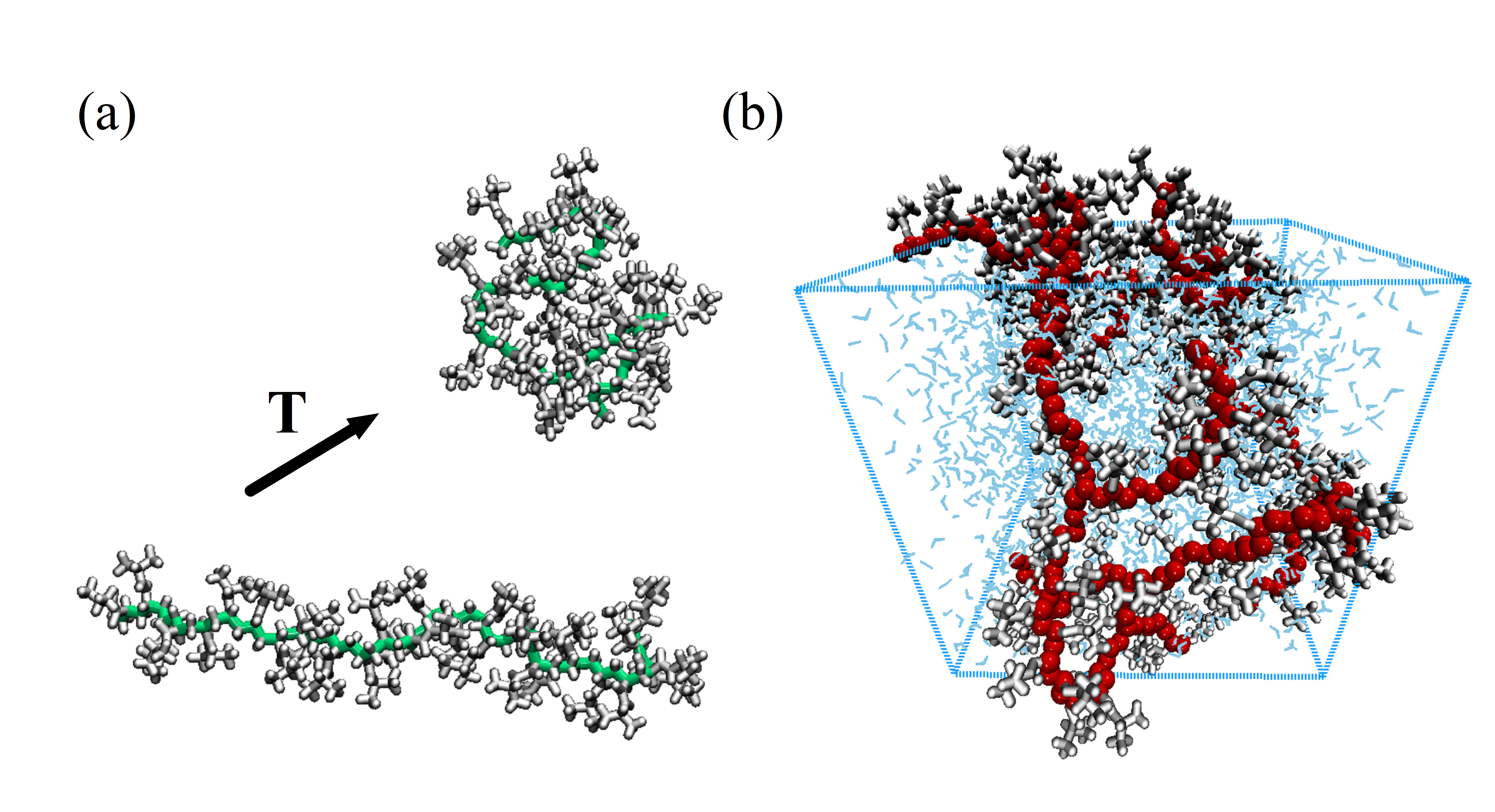}
\caption{\label{fig:atomistic}Snapshots from all-atom simulations representing (a) the coil-to-globule transition of a PNIPAM 30-mer in a dilute aqueous solution~\cite{tavagnacco2018molecular} (PNIPAM backbone heavy atoms and side chain atoms are displayed in green and white, respectively) and (b) a realistic model of a portion of a hydrated PNIPAM microgel particle~\cite{Zanattaeaat5895} (PNIPAM backbone heavy atoms, side chains atoms and water molecules are shown in red, white and blue, respectively).}
\end{figure*}

Despite the enormous increase of computational power, the task of performing all-atom simulations of entire microgels is still well out of reach. In general, the accuracy of atomistic simulations strongly depends on the quality of the model used to describe the interatomic forces acting between the atoms. When length- and time-scales of interest are too large to employ \textit{ab inito} methods, as in the case of PNIPAM, the choice of empirical force fields becomes particularly important. To explore PNIPAM properties, there exist several force fields that are able to reproduce the coil-to globule transition, such as AMBER~\cite{Du2010}, GROMOS~\cite{Walter2010, tonsing2001}, PCFF~\cite{deshmukh2012role} and OPLS~\cite{Walter2010,Tucker2012}. In addition, a modified version of the dihedral parameters in OPLS  provides an improved description of alkanes liquid properties~\cite{ Shiraga2015, Ester2016, tavagnacco2018molecular}.  Simulation studies that compare the capability of predicting the properties of PNIPAM by different force fields have been carried out both for the monomeric NIPAM unit~\cite{ Kamath2013} and for the oligomers~\cite{ Walter2010, Botan2016}. These comparisons have shown that the details of the conformational transition of PNIPAM depends on the specific pairs of force fields employed to describe PNIPAM and water~\cite{Walter2010}, and also that changes in the thermodynamic properties of the monomer affect the kinetics of the conformational transition~\cite{Kamath2013}.

In the context of microgels, high-resolution models can be very useful to understand and quantify processes that happen at the atomic and molecular length- and time-scales but whose effects extend well beyond them. For example, it is known that the volume-phase transition (VPT) exhibited by PNIPAM microgels is connected to the good-to-bad solvent transition that single PNIPAM chains experience in water at the lower critical solution temperature (LCST). Below the LCST the polymeric chains are in an extended conformation, while above this temperature they collapse into a globule state due to the complex interplay between PNIPAM-PNIPAM and PNIPAM-water interactions, which are temperature dependent, as schematically shown in Figure~\ref{fig:atomistic}(a). This phenomenon can be (and has been) carefully investigated with all-atom simulations. In particular, the high resolution of the atomistic models have helped evaluating the molecular parameters that can selectively tune the LCST value and therefore the corresponding VPT temperature (VPTT), with direct impact on the technological applications of microgels~\cite{doi:10.1021/acs.molpharmaceut.6b00942}. Indeed, atomistic simulations showed that the LCST increases by decreasing the degree of polymerization, thereby allowing to correlate the dependence of the transition temperature on the degree of polymerization to the effects of the chain length on the accessible conformations~\cite{Tucker2012}. Stereochemistry is another important molecular factor that influences the LCST value and that can be better understood with all-atom models. Simulation studies carried out on PNIPAM oligomers with different isotactic content revealed that the LCST value of PNIPAM chains with a high meso diad content, the isotactic PNIPAM, is lower than that of atactic oligomers because the diad composition affects the size, conformation and water affinity of the polymeric chain~\cite{Ester2016}. Below the LCST the isotactic stereoisomer prefers conformations with a lower radius of gyration and shows higher hydrophobicity. The effect on the LCST of another form of isomerism, such as the structural isomerism,  can also be tackled with an atomistic approach~\cite{Netz2018}. Interestingly, these detailed simulations demonstrate that polymeric chains with the same atomic compositions but different interconnections do not show the same LCST value, since hydrophobic interactions are affected by the spatial arrangement of the functional groups.

Atomistic simulations are also invaluable tools to gain insights into the molecular mechanism that drives the coil-to-globule transition~\cite{deshmukh2012role, tavagnacco2018molecular}. The chain transformation from an extended to a globule state was shown to occur with a relatively small loss of hydration water molecules and through a cooperative process that can be ascribable to the breaking of the hydrogen bonding network formed by water molecules in the proximity of hydrophobic groups. PNIPAM chains appear largely hydrated even above the LCST, and the coil-to-globule transition takes place with a significant rearrangement of the hydration water structure.

Atomistic modelling can be extended beyond single linear chains, as more complex molecular architectures can be represented and studied with all-atom simulations. This is the case, for example, of systems of crosslinked PNIPAM oligomers~\cite{tonsing2001}, polymeric membranes~\cite{adroher2017} or even portions of microgels~\cite{Zanattaeaat5895}. As shown in Figure~\ref{fig:atomistic}(b), it is now possible to simulate a cubic section (of linear size $\simeq 5~\mathrm{nm}$) of a microgel particle. These simulations have allowed to gather information on the temperature-dependence of its dynamical properties, finding a quantitative agreement with experimental results~\cite{Zanattaeaat5895}.

Another example of the application of atomistic modelling is provided by the phenomenon of co-non-solvency, which occurs when organic solvents such as short chain alcohols are added to aqueous solutions of PNIPAM. Even though water and alcohol are individually good solvents for PNIPAM, a mixture of the two cosolvents induces a collapse of the polymer for intermediate mixing ratios~\cite{wolf1978measured,tanaka_cononsolvency,hore2013co}. In the context of microgels, this counterintuitive phenomenon can be used to realize microgels that swell upon increasing the temperature above the VPTT~\cite{cononsolvency_pnipam_microgels}. For example, in the case of methanol-water mixtures it was shown that adding a small amount of methanol to a water solution of PNIPAM promotes a deswelling of the microgel, as PNIPAM and methanol molecules experience a favourable interaction that drives the collapse. By contrast, when methanol is added in excess the microgel re-swells in force of a favourable entropic contribution~\cite{water_methanol_pnipam}. Recently, detailed models have been used to quantitatively understand the origin of the co-non-solvency experienced by polymers dispersed in water/alcohol mixtures in both good and bad solvents~\cite{mukherji2017depleted}, providing an additional tool to control the responsiveness of microgels~\cite{klitzing_langmuir}.

In conclusion, it has been demonstrated that atomistic simulations can be very useful to describe the behaviour of small systems. However, they cannot be used to model the full complexity of multiple chains or polymers networks, which instead require the use of different approaches. For example, while all-atom force fields reproduce and provide a detailed description of the conformational transition of a single chain, extrapolating this information to understand the behaviour of a whole microgel is not straightforward~\cite{Botan2016}. Multiscale modelling techniques~\cite{Botan2016} and coarse-grained simulations can fill this gap and allow to quantitatively investigate the polymer phase behaviour.

\section{Monomer-resolved models}
\label{sec:monomers}

\begin{figure*}[t!]
\centering
\includegraphics[width=0.9\textwidth,clip]{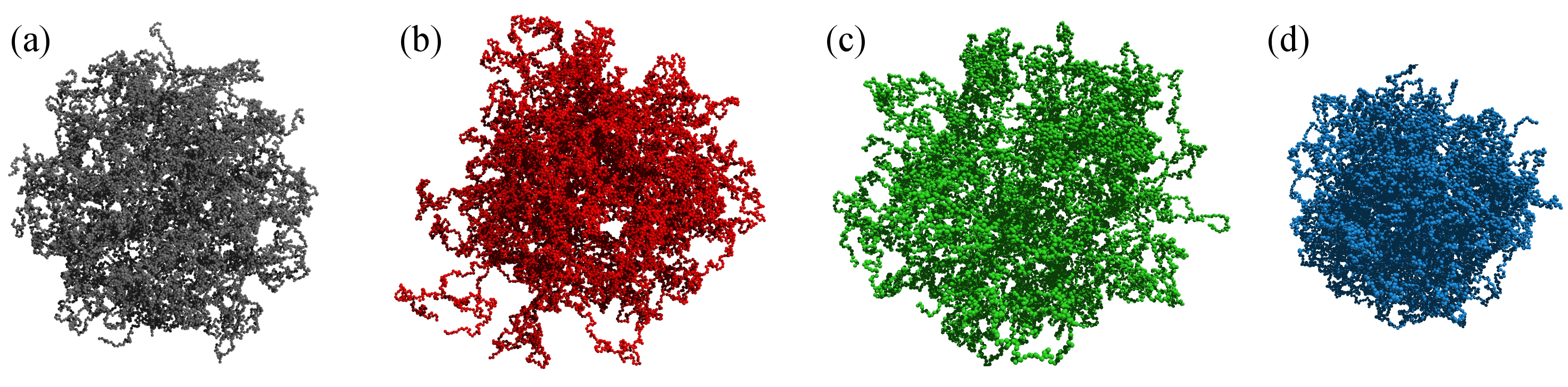}
\caption{\label{fig:snapshots}Snapshots of microgels generated with different protocols. (a) a diamond-lattice microgel, (b) a microgel assembled by pre-formed chains~\cite{moreno2018computational}, (c)-(d) microgels built from mixtures of patchy colloids assembled under two different concentration conditions~\cite{gnan2017silico}. }
\end{figure*}

In order to explore longer time- and larger length-scales, it is common to simulate complicated systems by using coarse-grained representations, which map groups of atoms or molecules onto single interaction sites~\cite{likos2001effective}. In the context of polymeric systems, the most prominent properties of polymers are known to be scale-invariant (at least in the limit of high polymerisation degree)~\cite{rubinstein2003polymer}, and hence rather insensitive to the microscopic (atomistic) details. This feature has been exploited to devise techniques that allow to systematically coarse-grain polymeric systems~\cite{capone_coarse_graining,d2012coarse,narros2014multi}, and chains are often modelled as collection of beads of size $\sigma_m$ connected by springs. In this representation, the size of the single beads is taken to be comparable with the Kuhn length of the real chain~\cite{rubinstein2010polymer}, which is often of the order of $\sim 1$~nm~\cite{rubinstein2003polymer,Fetters2007review}. For polymers in good-solvent conditions, the actual force-field used in MD simulations is usually the Kremer-Grest set of interactions~\cite{kremer1990dynamics}  which is the \textit{de-facto} gold standard. In this model the connectivity between bonded neighbours is provided by a finite-extension non linear elastic (FENE) spring, whereas the steric repulsion between all beads (bonded and non-bonded) is modelled through a Weeks-Chandler Andersen (WCA) interaction~\cite{WCA}; the parametrisation of these potentials makes sure that, under ordinary conditions of temperature and concentration, chains do not cross each other, so that the overall topology of the system is preserved. The Kremer-Grest model can be augmented by additional terms that can describe more specific cases, such as charged or semiflexible polymers~\cite{stevens1995nature,faller1999local, faller2000local}. 
The great majority of the numerical work done on polymer-based macromolecules deals with systems made of chains or aggregates with simple topologies such as rings~\cite{Halverson2011}, star polymers~\cite{Jusufi1999stars}, or dendrimers~\cite{Harreis2003} since, from the numerical point of view, the disordered nature of polymer networks poses significant challenges, in terms of both modelling complexity and computer time. Several strategies for the generation of suitable network topologies have been put forward, which are discussed and compared in depth in the next few paragraphs. A visual overview of some of the models discussed is presented in Fig~\ref{fig:snapshots}. The description of the different preparation protocols is completed by results on the swelling transition and on the swelling/deswelling kinetics of \textit{in silico} microgels.

\subsection{Protocols for the numerical design of coarse-grained microgel particles}

\subsubsection{Microgel formation from a crystalline lattice}
Early attempts of modelling  particles made by a crosslinked polymer network rely on placing the crosslinkers on a crystalline lattice (usually diamond) and connecting them with chains having the same size~\cite{claudio2009comparison,jha2011study,kobayashi2014structure}. The network is then cut-out  from a sphere to obtain the shape of the particle. Following the same approach it is possible to generate both standard and hollow microgels, which have been simulated to study the uptake and release of neutral species~\cite{schmid2016multi} or their behaviour at the liquid-liquid interface ~\cite{geisel2015hollow}. Recently, the diamond lattice procedure has  also been employed to generate microgels made of interpenetrated polymer networks~\cite{rudyak2018shell}. In addition,
the possibility of having chains of the same length represents an advantage when comparing numerical simulations with theoretical approaches. For instance, one can exploit scaling arguments based on the Flory theory for polymers or can apply the  blob tension model for evaluating the polymer stretching for different crosslinker concentrations as for instance done for ionic microgels~\cite{kobayashi2014structure, ghavami2016internal}. 

While this approach is suitable to showcase the potentialities of microgels for applications and for a direct test of the theory, it is far from being a realistic model. Indeed, lattice-based topologies suffer from several drawbacks: (i) all chains have the same length, something which is not found in real microgels; (ii) the distribution of crosslinkers is uniform by construction, while in the polymerization process employed to synthesize standard neutral microgels it is known that crosslinkers react faster than monomers, giving rise to an inhomogeneous distribution, with more crosslinkers in the core than in the outer corona~\cite{fernandez2011microgel,stieger2004small}; (iii) the corona is obtained from the spherical cut of the crystalline lattice and its extension can be controlled only by changing the chain length; (iv) the model has no loop-like defects and few or no dangling ends~\cite{claudio2009comparison}, which are fundamental for a comparison with real microgels, as they contribute to the hydrodynamic radius $R_h$ and might also play a role in the rheological properties of microgel suspensions at moderate concentrations~\cite{boon2017swelling}; (v) polymer chains do not entangle, which means that the elastic properties and the permeability of the diamond microgel depend on the length of the chains but not on the degree of entanglement of the polymers. Finally, the underlying crystalline structure of the diamond-lattice-based microgel affects the numerical density profiles and the form factors~\cite{gnan2017silico}, making a comparison with the experimental data difficult.

\subsubsection{Microgels from randomly distributed crosslinkers}
A step forward with respect to the previous protocol can be taken by randomly distributing crosslinkers within a cubic simulation box; close-by crosslinkers are then connected by polymer chains, for example by choosing a given cut-off distance~\cite{nikolov2018mesoscale}. This allows to generate non-ordered networks made of polymers chains that are slightly polydisperse. As for the crystalline-lattice-based microgels, the spherical shape is obtained from cutting out a sphere from the cubic simulation box. 
The main advantage of this method is that the crosslinker distribution can be fine-tuned, even though the connectivity among crosslinkers is not completely satisfied and thus cannot be fully controlled. We further note that this protocol makes it possible to generate core-shell~\cite{elvingson} or even hollow microgels~\cite{masoud2011controlled}, since the idea behind the assembly of the particle is similar to that employed for crystalline-lattice-based methods. 

\subsubsection{Microgels from the self-assembly of a gel network}
\label{sec:gel}
Taking inspiration from the synthesis process of PNIPAM microgels,  a recent numerical protocol has been developed to design coarse-grained microgel particles. The approach is based on the self-assembly of patchy particles, \textit{i.e.} hard-sphere particles decorated with attractive sites, which have shown to form gel networks at low and moderate densities~\cite{sciortino2011reversible}.
In Ref.~\cite{gnan2017silico} bivalent and tetravalent patchy particles are used to mimic, respectively, monomers and crosslinkers. Inter- and intra- species bonds are allowed except  for crosslinkers particles that cannot form bonds among themselves. Although the interactions among monomers and crosslinkers resemble those occurring in the synthesis of real microgels, the dynamic processes that lead to the network formation in the two cases have little in common. Indeed, while the polymerization mechanism in real microgels is an off-equilibrium process, the build-up of the network by the patchy particles occurs in equilibrium, and is facilitated by a 
``swapping'' mechanism~\cite{Sciortino2017} which allows the system to  easily equilibrate even at low temperatures, where the fully-bonded-network condition can be almost accessed. The self-assembly process allows to generate a disordered network where the length of the polymer chains follows an exponential distribution which can be predicted by a heuristic argument based on the Flory theory in the fully bonded limit~\cite{rovigatti2017internal}. Instead of cutting out a spherical region from a bulk homogeneous network, a spherical confinement (mimicking confinement in a droplet) is employed during the self-assembly process. Such external field acts as an extra parameter which can be tuned to influence the topology of the network: by varying the radius of the spherical confinement  at fixed crosslinkers concentration it is possible to generate microgel particles with different density, topology and degree of entanglement. The resulting microgels can range from compact to floppy with several dangling ends, experiencing very different swelling behaviors~\cite{gnan2017silico}. As a result, this protocol makes it possible to investigate the role of topology on the dynamics of swelling in a unique way.  As for the case of crystalline-lattice-based microgels, the resulting network possesses a homogeneous distribution of crosslinkers. However, the corona spontaneously arises from the interfacial region formed by the system due to the presence of a confining field; with this approach the width of the corona can be controlled by the thermodynamic properties of the network, \textit{i.e.} by temperature and density.

\subsubsection{Microgels from the  assembly of functionalized chains}
In addition to the standard synthesis process  based on precipitation polymerization, whereby the monomer and the initiator form an homogeneous phase whereas the obtained polymer is insoluble and precipitates~\cite{fernandez2011microgel}, it is also possible to synthesize microgels through microfluidics fabrication using droplets of macromolecular precursor chains that are later photo-crosslinked~\cite{seiffert2010controlled}. Inspired by this technique, Moreno and Lo Verso devised a new numerical protocol for assembling microgel particles that exploits the self-assembly of pre-formed chains which are functionalised with reactive groups~\cite{moreno2018computational} and placed in a spherical confinement. In the method presented in Ref.~\cite{moreno2018computational}, a fraction $f$ of reactive groups are placed randomly on the polymer chain, with the constraint that consecutive reactive sites are not allowed in the backbone sequence. During the dynamics reactive sites form permanent bonds; this gives rise to a  fast stage in which the majority of reactive groups bond together, followed by a second slow stage in which non-bonded reactive sites seek out other reactive groups until full crosslinking is achieved. The latter stage  is sped up by selecting randomly two non reacted sites and by applying an attractive external field between the two that allows them to get in contact and form a bond. Differently from crystalline-lattice-based microgels, this procedure allows to design microgels with entangled polymer chains of different sizes, together with a conformational polydispersity which is fundamental for investigating the role of topology in the swelling dynamics of the particle. In addition, the number of the precursor chains is independent on the number of crosslinkers, which allows to prepare microgels with different densities at fixed crosslinker concentrations. Finally the presence of a spherical confinement provides a spherical shape to the assembled-network  without the need to cut it out from the bulk of the polymer network. A similar procedure was employed in Ref.~\cite{minina2018influence}  where the reactive sites are not chosen at the beginning of the simulation, but at the end. Namely, a number of polymer chains are equilibrated within a confined network and then crosslinked by selecting randomly monomers among those separated by a maximum distance. Unlike in the procedure of Ref.~\cite{moreno2018computational}, only the crosslinking of sites belonging to different chains is allowed, which artificially suppresses the formation of intrachain loops. If the final crosslinker concentration is smaller than the one desired, the maximum bonding distance is increased and the procedure  is repeated. Although similar to the technique described previously in this section, this strategy gives rise to non-compact microgels even when a high crosslinkers concentration is chosen.

\subsubsection{Comparison of the assembly protocols: topology and form factors in the swollen state}

\begin{figure}[h!]
\centering
\includegraphics[width=0.5\textwidth]{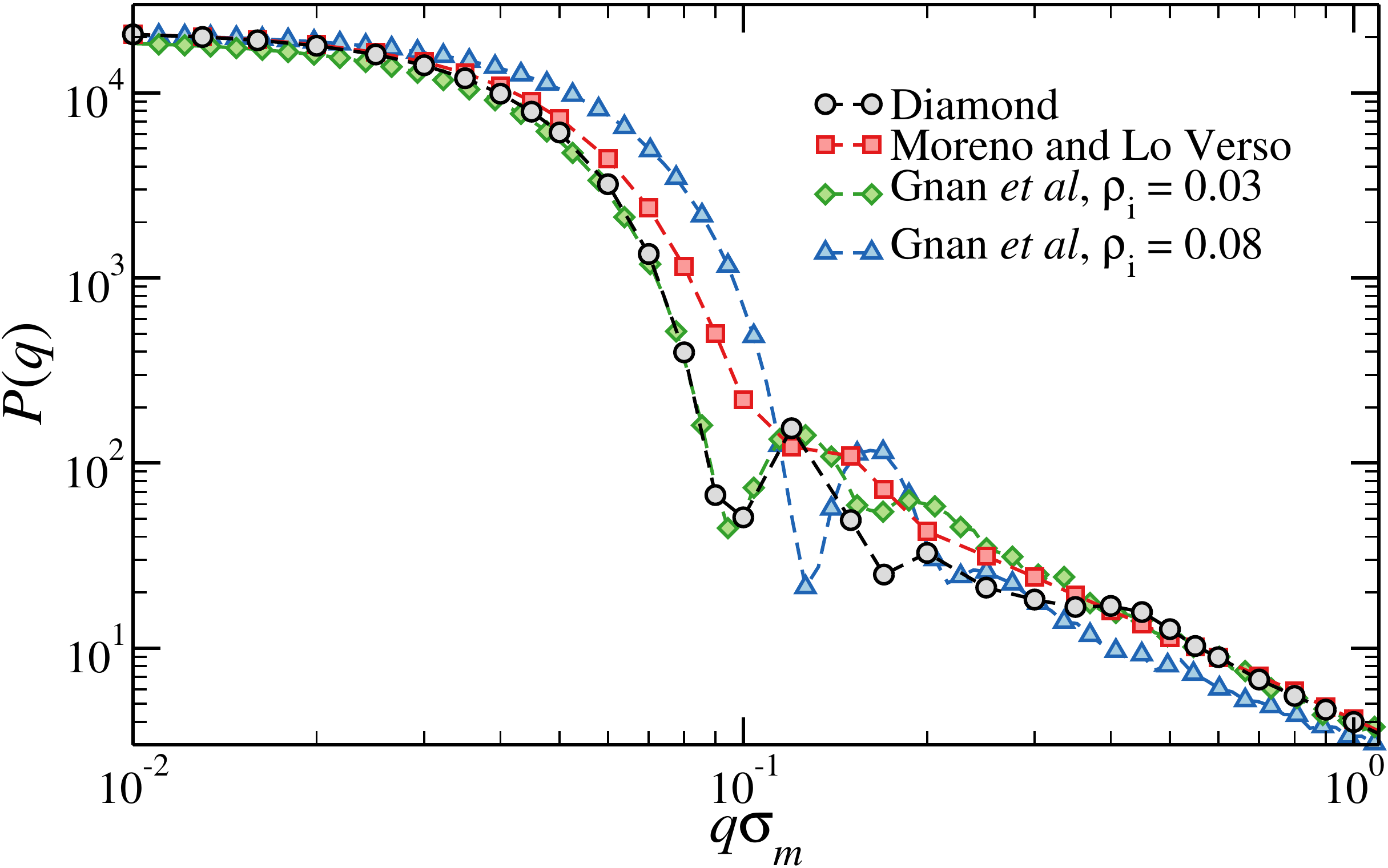} 
\caption{\label{fig:pq_comparison}The form factor of microgels generated with different numerical protocols across the volume-phase transition~\cite{gnan2017silico,moreno2018computational}. The two microgels assembled with the protocol of Ref.~\cite{gnan2017silico} have been generated within two different spherical confinements, here indicated by the overall density of the mixture $\rho_i$.}
\end{figure}

We conclude the overview of the protocols used to build different topologies by comparing some of the microscopic architectures discussed.
Figure~\ref{fig:snapshots} shows representative snapshots of the diamond microgel and of microgels generated with the methods of Refs.~\cite{moreno2018computational} and~\cite{gnan2017silico}.

Figure~\ref{fig:pq_comparison} shows the form factor, $P(q)$, where $q$ is the wavevector, of microgels composed of $N \approx 21000$ monomers and a crosslinker concentration of $\approx 1.2 \%$. The form factor of the diamond-lattice microgel, which we use here as a reference, displays a peak at a position that roughly corresponds, in real space, to the size of the particle. At larger wavevectors ($0.2 \leq q\sigma_m \leq 0.4$), the form factor exhibits a weak dependence on $q$ that is due to the underlining ordered structure of the network and can therefore be considered spurious~\cite{gnan2017silico}. The other curves refer to disordered topologies, either generated with the method by Moreno and Lo Verso~\cite{moreno2018computational}, or through the assembly of binary mixtures of patchy particles~\cite{gnan2017silico}. In all cases we observe a peak or a shoulder around $q\sigma_m \approx 0.15$ that is linked to the size of the microgels and, for $q\sigma_m \geq 0.3$, very similar decays, reflecting the self-avoiding character of the strands~\cite{rubinstein2003polymer}. Contrarily to the diamond case, the form factors of each disordered topology are compatible with the fuzzy-sphere model~\cite{fernandez2011microgel}. The difference between the different topologies is concentrated in the intermediate $q$-region: microgels generated at higher densities display a more structured $P(q)$. The lack of well-resolved peaks in the form factor of the microgel assembled with functionalized chains~\cite{moreno2018computational} signals the larger heterogeneity of the network compared to the case of microgels generated by assembling patchy mixtures.

\subsection{Swelling and solvent effects}
After assembling the network, the next relevant issue to address is reproducing the swelling/deswelling transition of microgels. To this aim, the solvent effects must be taken into account, and this can be done at various levels of coarse-graining. Apart from the atomistic route discussed in Section~\ref{sec:atomistic}, even for a single microgel the simulations in the presence of a coarse-grained solvent can be computationally expensive. Thus, it is preferable to use implicit solvent models to reproduce those features of the volume phase transition that are independent of the actual presence of the solvent, such as the thermodynamic and geometrical properties across the VPT, and then to resort to explicit solvent models to tackle specific problems for which the presence of the solvent is absolutely needed, such as for example solvent expulsion, kinetic aspects and interfacial problems.

We now start to discuss the so-called implicit solvent models, where an effective `solvophobic' potential between the monomers which takes into account the affinity between polymer and solvent is introduced. The solvophobicity is normally modulated by a control parameter $\alpha$ which plays the role of the temperature (or of the external parameter controlling the swelling). From the practical standpoint, the effective potential is a monomer-monomer interaction that is non-negative under good solvent (or maximally swollen) conditions, and becomes very attractive under bad solvent (or collapsed) conditions. To this purpose, the use of a simple Lennard-Jones potential may give rise to unphysical non-monotonic behavior of the microgel size~\cite{keidel2018time} with increasing quench depth due to the relative contribution of the attraction and the repulsion when the potential depth increases. Instead, the potential  initially proposed by Soddemann and coworkers~\cite{soddemann2001generic} was found to  well reproduce the swelling behavior for microgels assembled in different ways~\cite{gnan2017silico,moreno2018computational}.  We further note that the specific form of this solvophobic interaction is not expected to play a major role in the swelling behavior, at least from the qualitative point of view, and thus different choices could be adopted, as for example the potential used to model star polymers~\cite{huissmann2009star} or telechelic star polymers\cite{rovigatti2016soft} in solvents of different quality.

\begin{figure}[h!]
\centering
\includegraphics[width=0.5\textwidth]{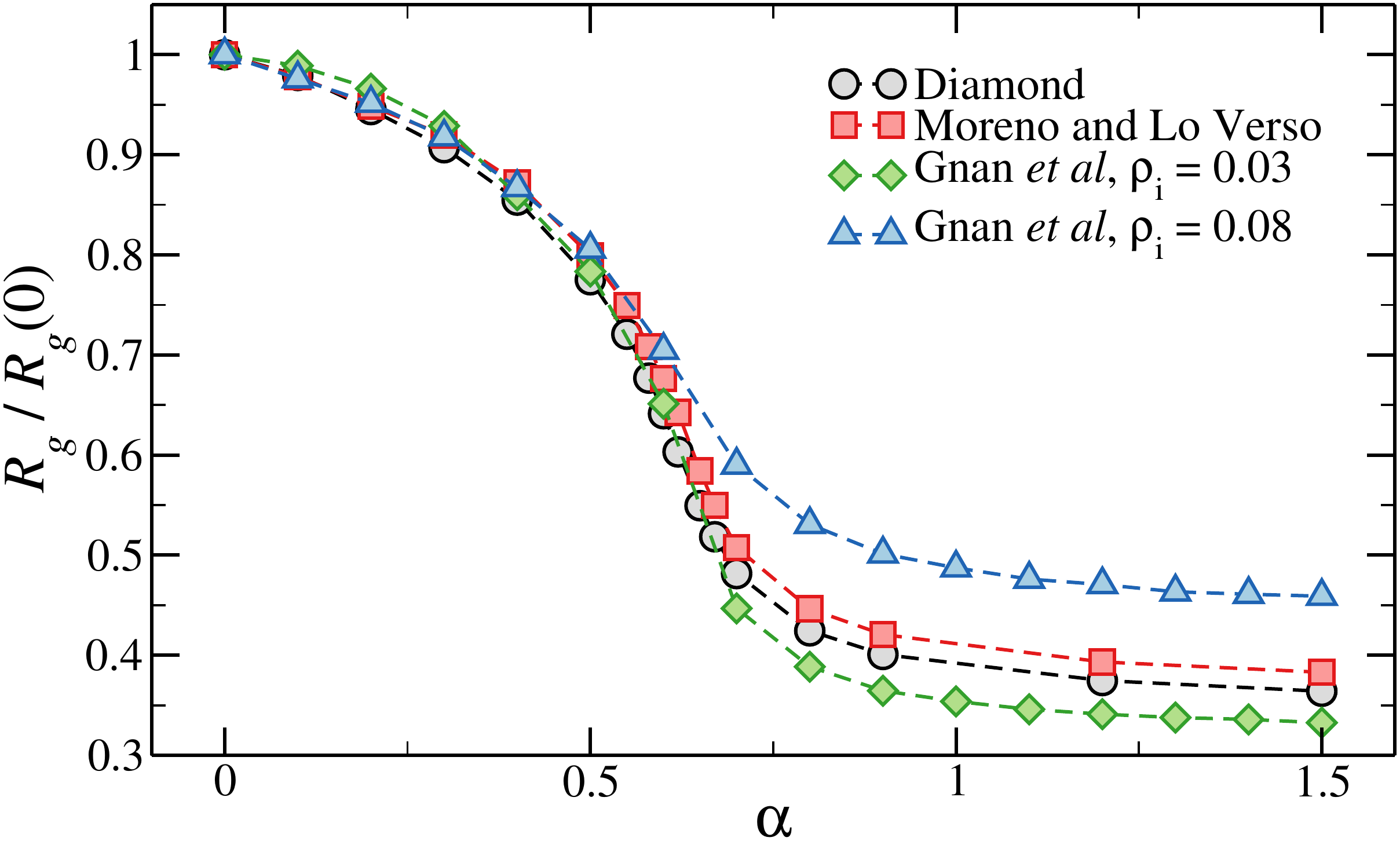} 
\caption{\label{fig:swelling}The relative change of gyration radius of microgels generated with different numerical protocols across the volume-phase transition. The two microgels assembled with the protocol of Ref.~\cite{gnan2017silico} have been generated within two different spherical confinements, here indicated by the overall density of the mixture $\rho_i$. Here $\alpha$ is the parameter of the monomer-monomer interaction that controls the quality of the solvent, playing the role of the temperature in real PNIPAM microgels.}
\end{figure}

Interestingly, Fig.~\ref{fig:swelling} shows that the swelling curve is not very sensitive to the inner topology of the network. Indeed, microgels generated in different ways at approximately the same crosslinker concentration, including also the diamond lattice microgel, display very similar swelling properties and even the same VPTT~\cite{moreno2018computational}. Of course, with an underlying ordered lattice, the only way to tune the topology is to vary the chain length, affecting the crosslinker concentration. Instead, using disordered assemblies, the variation of the confining volume can be used to significantly alter the swelling properties\cite{gnan2017silico,rovigatti2017internal}.

The incorporation of an explicit solvent is the next step of description. Of course, this cannot be done at the atomistic level of accuracy, but still sometimes  simple potentials like Lennard-Jones or modifications there-of have been adopted. These have the main disadvantages that excluded volume of the solvent molecules can be sometimes overestimated\cite{camerin2018modelling}. Thus, it is much better to rely on a coarse-grained solvent representation, where groups of solvent molecules are treated as soft beads. This is precisely the aim of the Dissipative Particle Dynamics (DPD) technique, which has the advantage of correctly reproducing hydrodynamics at long times by imposing locally the conservation of momentum\cite{espanol1995statistical}. In addition, the DPD method has been mapped to polymer-solvent interactions and provides a way to directly relate the parameters of the involved soft potentials to the Flory-Huggins solvency parameter\cite{groot1997dissipative}. However, in order to do so, it is necessary to use such soft potentials among all species involved, including monomer-monomer interactions. This may give rise to unphysical crossing between polymer chains and care must be taken when adopting this method~\cite{rudyak2018shell}.

DPD has been used in a number of studies\cite{gumerov2016mixing,Rumyantsev2016,Mourran2016} performed with a regular diamond network.  It was also used by Nikolov and coworkers for a topology obtained by using randomly distributing crosslinkers~\cite{nikolov2018mesoscale}. However, in order to compare the effect of the explicit solvent with the widely used implicit ones, a one-to-one correspondence must be established. This was the aim of a recent work~\cite{camerin2018modelling} where identical microgel configurations, interacting through the usual Kremer-Grest force field, were compared in implicit and explicit solvent conditions for both swelling curves and form factors. It was shown that a DPD treatment of the solvent gives a faithful representation of the implicit solvophobic potential in all aspects, opening the way for a systematic use of the explicit solvent to investigate interfacial aspects of microgels. For example, an interesting aspect to model is the flattening of the soft colloids, and particularly microgels with their inhomogeneous core-corona structure, at a liquid-liquid interface\cite{style2015adsorption}, that is relevant for applications as emulsion stabilizers\cite{geisel2012unraveling}.

Finally, some studies have also adopted the more accurate Multi-Particle-Collision-Dynamics to treat the solvent~\cite{ghavami2016internal,keidel2018time}, but only for the diamond-lattice topology. These simulations have shown that the monomer dynamics under swollen conditions agree with the predictions of the Zimm model~\cite{rubinstein2003polymer,Doi1986}. However, as the microgel shrinks by decreasing the solvent quality, the dynamics progressively deviates from this theoretical model and, in the fully collapsed state, hydrodynamic interactions are screened out while the dynamics approaches the predictions of the Rouse model expected for polymer melts~\cite{rubinstein2003polymer,Doi1986}.

\subsection{Kinetics of swelling and deswelling}

Monomer-resolved simulations make it possible to study in detail the evolution of the microgel internal structure under changing the solvent conditions --- a feature that is not easily accessible in experiments --- and to unravel the effect of the network microstructure on the kinetics of swelling and deswelling. The time evolution of the microgel radius of gyration during its collapse, $R_{\rm g}(t)$,  was analyzed in
Refs.~\cite{elvingson,moreno2018computational,camerin2018modelling,keidel2018time} for diamond and disordered microgels.
Ref.~\cite{nikolov2018mesoscale} also analyzed the swelling, finding consistency with Tanaka's theory \cite{nikolov2018mesoscale}.
Some general trends were observed for all models of microgels. In particular, higher degrees of crosslinking \cite{nikolov2018mesoscale}, higher regularities of the network \cite{moreno2018computational} and deeper quenches (to poorer solvent conditions) \cite{moreno2018computational} all result in faster and less stretched decays of $R_{\rm g} (t)$. Very interestingly, the shape of $R_{\rm g} (t)$ is apparently independent of the solvent model \cite{camerin2018modelling}. 

\begin{figure}[h!]
\centering
\includegraphics[width=0.5\textwidth]{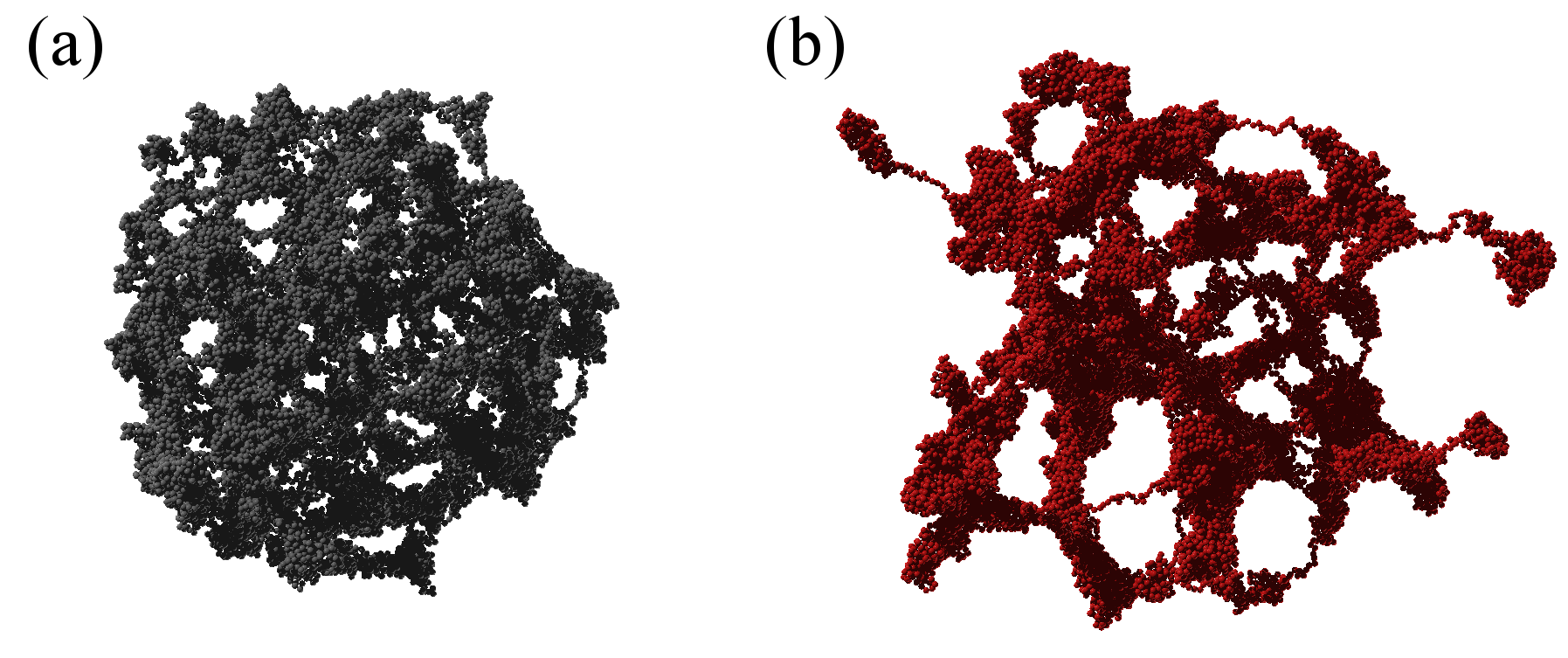}\\
\includegraphics[width=0.5\textwidth]{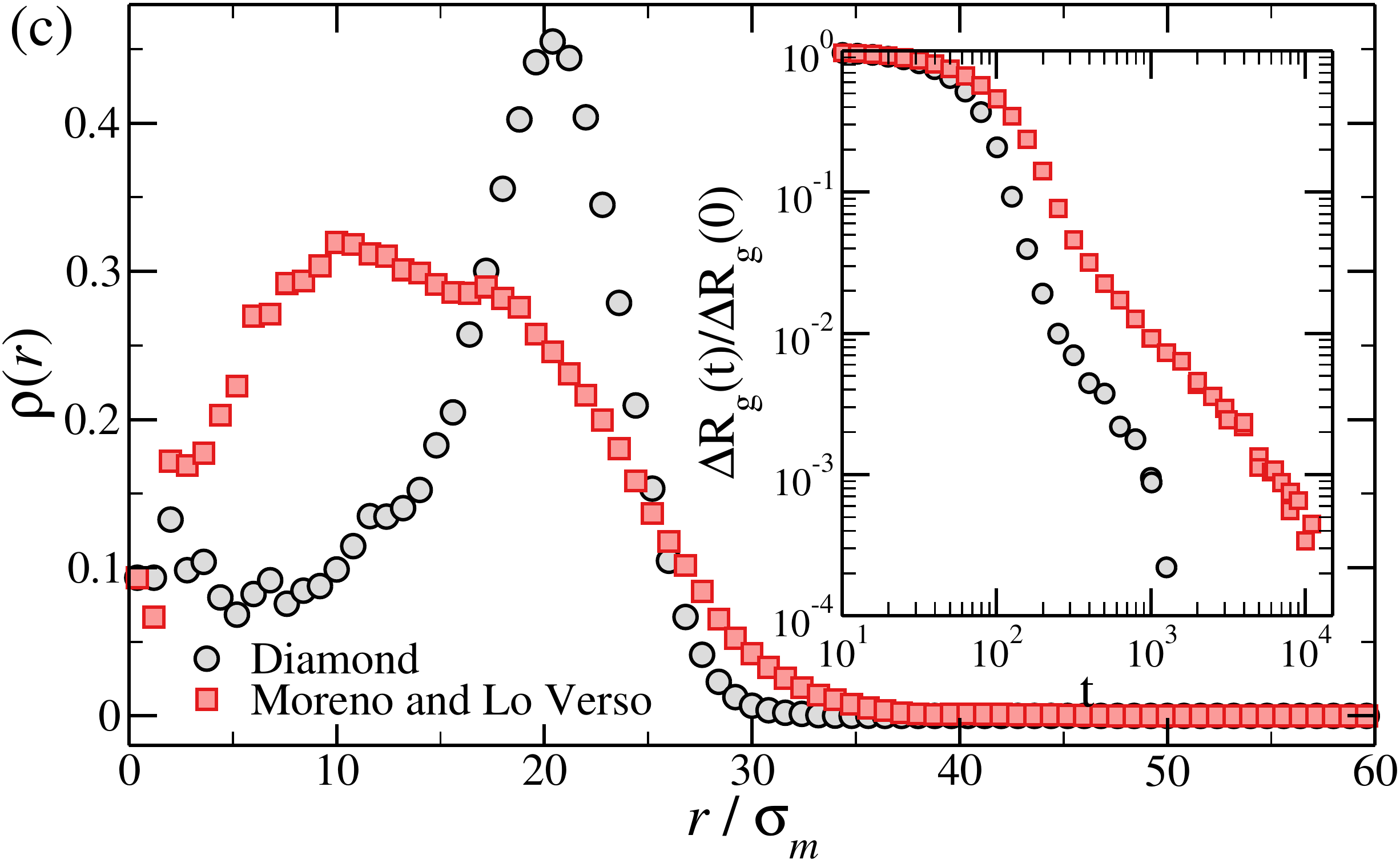}
\caption{\label{fig:coarsen} Top row: snapshots of (a) a diamond network and (b) a disordered microgel obtained by crosslinking polymers \cite{moreno2018computational} at intermediate times of the deswelling. Bottom panel (c): the density profiles of the same microgels calculated during the coarsening, namely at the same relative shrinking $\Delta R_{\rm g}(t)/\Delta R_{\rm g}(0) = 0.2$, with $\Delta R_{\rm g}(t) = R_{\rm g}(t)-R_{\rm g}(\infty)$. The inset shows $\Delta R_{\rm g}(t)/\Delta R_{\rm g}(0)$ for the same microgels.}
\end{figure}

As the microgel collapses when it is driven beyond the VPT point,
the monomers start to form  local globules that  progressively merge into interconnected larger domains,
until the whole structure is finally joined into a single dense spherical globule.
This phenomenon is known as 'coarsening' and is universally observed in phase separating systems \cite{Durian1991,Lambert2010,Bray2002}.
Direct visual inspection of simulation snapshots at intermediate times of the coarsening
reveals rather different conformations depending on the topology of the network.
Both regular and disordered networks with relatively high degree of crosslinking
are approximately spherical at all times, from the initial swollen to the final collapsed state \cite{nikolov2018mesoscale,moreno2018computational}.
Instead, disordered ones with low degree of crosslinking display
at intermediate times irregular conformations, with significant asphericity \cite{elvingson,moreno2018computational}
and large globulated protrusions \cite{moreno2018computational,camerin2018modelling}.
Two illustrative examples of these conformations are presented in Fig.~\ref{fig:coarsen}(a)-(b).

The homogeneous or heterogeneous character of the collapse of the outer shell is also reflected in the
monomer density profiles calculated with respect to the center of mass. In regular networks or in dense disordered ones,
the initial stage of the coarsening is characterized by a strong monomer aggregation in the outer shell, while the core of the
microgel remains 'hollow' \cite{keidel2018time}. This effect is much less pronounced 
in the heterogeneous collapse of disordered low-density microgels
(see main panel of Fig.~\ref{fig:coarsen}(c)), for which a flat density profile in the core is quickly reached. We also mention that the time-dependent size of the microgels during the collapse depends crucially on the topology, as exemplified by the inset of Fig.~\ref{fig:coarsen}(c).

The coarsening kinetics of the microgel deswelling has been characterized in Ref.~\cite{moreno2018computational}
by measuring the length of the growing domains. A clear difference was found between the regular diamond networks and the disordered microgels constructed by crosslinking of functionalized chains. The latter show a  power-law time dependence for the growing domain length, which has been suggested \cite{moreno2018computational} to be an intermediate scenario between liquid-gas phase separation \cite{Testard2014} and collapse of linear chains \cite{MajumderEPL2011}.
Though a similar power-law domain growth is observed at early times in the diamond networks, an accelerated growth  is  found at the late stage of the coarsening \cite{moreno2018computational}.  The observed master functions for the domain growth in both kinds of microgels are independent of the depth of the quench (\textit{i.e.}, of the solvent quality parameter)~\cite{moreno2018computational}. A similar result has been found in Ref.~\cite{elvingson}.
Remarkably, in analogy with general observations for critical phenomena, a scaling relation between dynamic correlators and the growing domain size has been found \cite{moreno2018computational}, with the scaling function being independent of the network microstructure.   
Finally, it is worth mentioning that, although no quantitative analysis has been reported, the conformations presented as simulation snapshots in Ref.~\cite{nikolov2018mesoscale} display no significant coarsening in {\it swelling} microgels, suggesting
a much more homogeneous character of the expansion of the monomers starting from the collapsed globular state. 
	
\section{Further coarse-graining}
\label{sec:coarse-graining}

Microgel suspensions have become a model system in fundamental physics, allowing to shed light on diverse phenomena such as jamming~\cite{zhang2009thermal,conley2017jamming} and glass~\cite{mattsson2009soft} transitions, charge effects~\cite{fernandez2000charge,bysell2010effect}, depletion interactions~\cite{rossi2015shape,maxime_nat_comm}, and more~\cite{fernandez2011microgel}. In this context, the complex internal architecture and the resulting responsiveness of the single microgels are crucial ingredients that can be harnessed to steer the collective behaviour of the system. However, a numerical description of a bulk system that make use of monomer-resolved models, which include these features by construction, is out of reach. Indeed, such a detailed description would require the total number of degrees of freedom to be so large that accessing the length- and time-scales that are characteristic of the phenomena of interest would be impossible with modern-day numerical resources. State-of-the-art simulation tools only allow to look at the static behaviour of dense systems made of multiple microgels, for example to investigate the structural changes of single microgels upon compression in overcrowded environments~\cite{scotti_hollow}, as long as the phenomena of interest do not require microgel diffusion. The obvious solution is to employ much simpler models that contain the minimal number of ingredients required to observe the desired bulk behaviour.

A simple approximation is to describe microgels as spheres that interact through a soft potential, which usually takes the form
\begin{equation}
\label{eq:vr}
V(r) = U_0 \left(1 - \frac{r}{\sigma}\right)^n,
\end{equation}
\noindent
where $U_0$ is a prefactor linked to the overall softness of the interaction, $\sigma$ is an effective particle diameter and $n$ is commonly set to 2 (harmonic potential) or $5/2$ (Hertzian potential). The latter choice can be theoretically justified by leveraging the classical elasticity theory (CET)~\cite{landau1986theory}. Indeed, in the CET framework two elastic spheres in contact experience an Hertzian effective repulsion. The CET assumes that the two objects are homogeneous and is strictly valid only in the small deformation regime, \textit{i.e.} when the centre-to-centre distance between the two particles is not too small. 
In early works, microgels were compared to hard sphere behavior, particularly for the dependence of the zero-shear viscosity on packing fraction $\zeta$ and found that up to about $\zeta \approx 0.5$, no significant differences were observed\cite{senff1999temperature}. However, above this threshold, while the hard sphere viscosity would diverge close to $\zeta \approx 0.58$, data measured for microgels show a clear deviation. The rheological data of elastic moduli were found to obey a power law increase that would be compatible with a soft sphere potential, particularly of an inverse power law form with exponent between 9 and 12.  However, later on, experimental evidence based on microscopy measurements in dilute conditions lent support to an effective Hertzian potential~\cite{zhang2009thermal}. These findings were confirmed by quantitative comparisons based on confocal microscopy experiments in the fluid phase and simulations which showed that, treating the microgels as elastic Hertzian spheres, a good description of the radial distribution functions across the whole fluid concentration region is obtained\cite{mohanty2014effective}.

\begin{figure*}[t!]
\centering
\includegraphics[width=0.9\textwidth,clip]{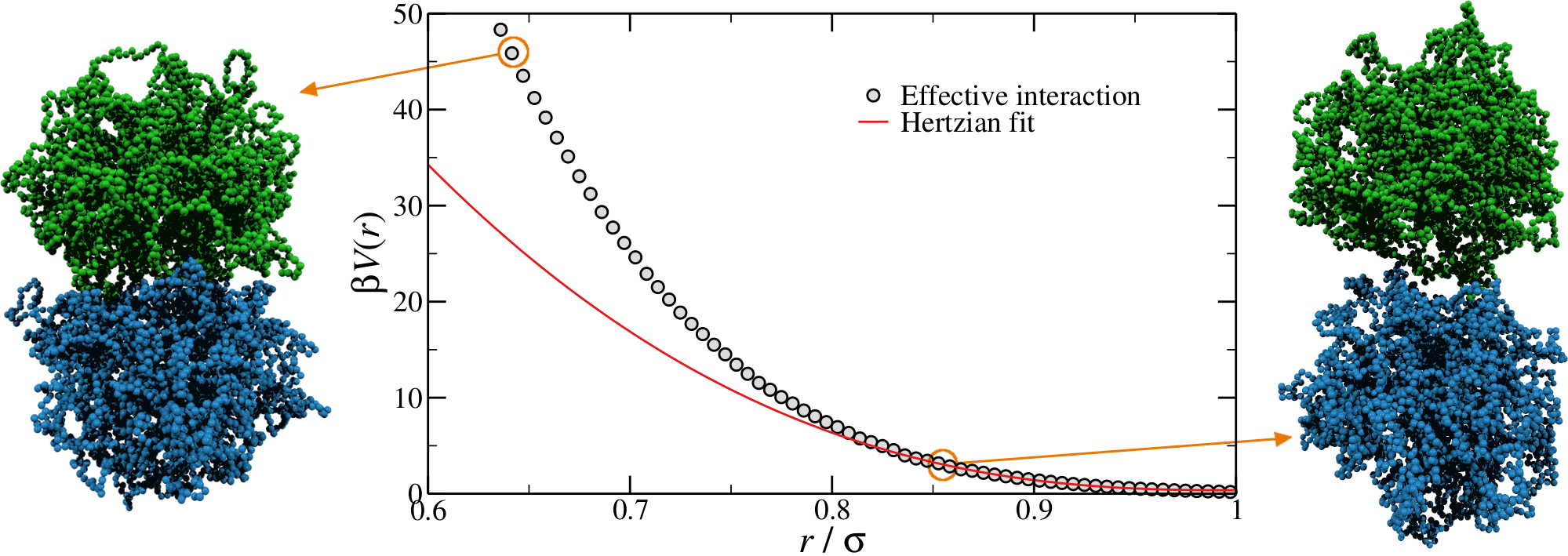}
\caption{\label{fig:hertzian}The effective interaction between two microgels with $5\%$ crosslinker concentration, as computed with monomer-resolved numerical calculations (points) and as estimated by using the simple Hertzian form of Eq.~\eqref{eq:vr} (line)~\cite{rovigatti2018validity}. The snapshots on the left and on the right show two representative conformations at large and small separations $r$, as indicated by the circles. Here $\beta = 1 / k_B T$, where $k_B$ is the Boltzmann constant.}
\end{figure*}

In order to provide a numerical test of the CET assumptions and of the range of validity of the Hertzian model, the calculation of the effective potential between two microgels is required. Such a study was done  for two diamond-lattice-based small microgels  by Ahualli and coworkers~\cite{Ahuali2017}, who reported that the resulting effective potential was not compatible with the Hertzian model, but rather with a generalized form of it, where the exponent $n$ in Eq.~\ref{eq:vr} varies, being best described by the value 3.5. However, this description is purely phenomenological and not based on any elasticity theory concept. Going one step forward and really verifying the Hertzian model would require the calculation of the elastic moduli of individual microgels, entering in the prefactor of Eq.~\ref{eq:vr} for $n=5/2$, namely $U_0=(2 Y \sigma^3)/(15(1 - \nu^2))$ where $Y$ is the Young modulus and $\nu$ the Poisson's ratio.  Using microgels with a disordered topology, generated from the self-assembly of a gel network with the patchy mixture method described in Section~\ref{sec:gel}, recent simulations provided the numerical evaluation of all elastic moduli as well as of the microgel-microgel effective interaction. This work, differently from ~\cite{Ahuali2017}, confirmed that the effective potential is well described by the Hertzian model at small deformations or, equivalently, for repulsion energies of a few times the thermal energy $K_B T$~\cite{rovigatti2018validity}. Above these limits, strong deviations are found, as shown in Fig.~\ref{fig:hertzian}, where it is evident that the Hertzian picture breaks down when  particles come into close contact, becoming anisotropic. These results have also been  quantified in a packing fraction range of validity of the Hertzian model, by means of bulk simulations with the Hertzian model and with the numerically calculated potential, yielding good agreement between the two up to packing fractions $\zeta \sim 0.8-1.0$, depending on the crosslinker concentration~\cite{rovigatti2018validity}. These findings confirm and extend earlier results in the fluid regime, but cast relevant doubts about the use of the Hertzian interaction to describe dense microgel suspensions that undergo glass or jamming transitions, a procedure that is largely employed in the literature\cite{zhang2009thermal,yang2011glassy2,seth2011micromechanical}. The non-Hertzianity of microgel-microgel interactions has also been recently quantified, and a multi-Hertzian description (which models the effective interaction as a sum of Hertzian contributions) has been shown to yield good agreement with experiments~\cite{maxime_nat_comm}. We note on passing that, as microgels approach the close-packing state, they experience a deswelling due to the steric compression~\cite{romeo_deswelling} that may further alter their mutual interaction.

At high packing fractions, not only the Hertzian picture breaks down, but also (and probably most severely) the two-body approximation. Thus many-body effects must be taken into account. This is currently a rapidly evolving field of research activity. Among the early attempts to go beyond the Hertzian model, with a simple but effective modification is the numerical work of Urich and Denton~\cite{urich2016swelling}, introducing the possibility of changing the particle size isotropically. A similar method has been used more recently to investigate how softness affects the dynamics of dense suspensions~\cite{higler2018apparent}. This approach has also been extended to charged microgels~\cite{weyer2018concentration}, showing that the model is able to capture the swelling behavior as a function of concentration observed in experiments~\cite{pelaez2015impact}. A step forward would be to also allow shape changes, as done in an early work related to crystallization aspects only~\cite{batista2010crystallization}.

Several recent works have also accounted for osmotic compression and deswelling to compare with experimental systems within simple models. In particular, van der Scheer and coworkers provide a phenomenological description of the internal equation-of-state of soft particles to predict the behavior of the collective relaxation~\cite{van2017fragility}, while de Aguiar {\it et al}  employ a modification of the Flory-Rehner~\cite{flory1953principles} theory which explicitly includes the stretching of the chains~\cite{de2017deswelling}. 
All these efforts represent concrete steps forward with respect to simple pair-wise models, but additional work will be needed in the near future to develop a more microscopic model which takes into account the internal degrees of freedom of realistic soft particles, thus being able to describe interpenetration~\cite{mohanty2017interpenetration}, deformation and faceting~\cite{de2017deswelling,conley2017jamming}. To consider these aspects in more refined micromechanical models~\cite{bonnecaze2010micromechanics} will be an important issue for a more quantitative prediction of rheological properties of dense and jammed states.

\section{Perspectives}
\label{sec:perspectives}

There are several directions towards which future work aiming at providing a numerical description of microgels should proceed.
Regarding atomistic simulations, there are a few aspects that could be explored in order to better understand the structural and dynamical properties of polymer-solvent interactions. In particular, more complex topologies, such as polymeric networks of different degree of disorder, could be exploited to investigate the molecular behaviour across the VPT and to evaluate the dominant interactions throughout the process. Another possibility, that takes advantage of the explicit description of the solvent at the atomistic level, is to use different architectures to better understand the interplay between water and polymer and to characterize to what extent the solvent affects the microgel properties. In this context, a very promising use of these atomistic approaches would be to quantify the interactions arising between polymeric oligomers at the molecular level and incorporate them in more coarse-grained models.
 
Concerning the monomer-resolved approaches that have been proposed so far to produce disordered networks, we also foresee several strategies for improving the current understanding. Starting from the protocol based on patchy particles self-assembly, it is important to stress that, while this was not designed in order to reproduce realistically any experimental synthesis process, it aims to achieve a fine control of the resulting assembly product. To achieve this goal, it appears extremely important to be able to control the distribution of crosslinkers inside the network, in order to produce more and more realistic network topologies which would closely resemble experimental measurements, \textit{e.g.} by recent super-resolution microscopy\cite{bergmann2018super}. In parallel, such a control would open up the possibility to tailor the resulting density profiles at will. Since the assembly of the patchy mixture occurs in equilibrium, such a finer degree of control might be obtained by either changing the force field or by adding external fields. Another possible extension of the protocol presented in Ref.~\cite{gnan2017silico} would be to change the numerical procedure so as to explicitly reflect the experimental synthesis at the microscopic level. For example, since polymerization is a one-directional process, in a realistic numerical synthesis the chains would start growing from a small fraction of one-patch particles, which would play the role of the initiators of the polymerization process. The free monomers, or the end monomers of a growing chain, would then only react with the end monomers of other growing chains. Likewise, branching would only occur at the ends of the growing chains. Another ingredient that might be incorporated in the model is the difference between the reaction rates of chain growth and branching. Thus, the fraction of initiators and the rates for chain growth and branching would act  as additional control parameters, which can be adjusted to control the topology of the resulting network.

About the numerical protocol based on crosslinking of pre-existing chains~\cite{moreno2018computational}, it already reflects the experimental synthesis from a microfluidic approach in a reasonably realistic fashion. However, possible improvements of this method should take into account the directionality of the interactions between the reactive groups, thereby penalizing the formation of intra-chain small loops. In addition, a question to address in the future is how the mechanisms implemented to accelerate the synthesis, such as random crosslinking of the last unreacted groups in~\cite{moreno2018computational} but also bond swapping in~\cite{gnan2017silico}, may affect the resulting microstructure. For instance, random crosslinking of the last groups makes the non-trivial assumption that the energy barriers that impede the formation of the few remaining unbound pairs can be overcome by waiting long enough. On the other hand, bond swapping may prevent the freezing of entropically unfavourable local structures that would emerge in a purely irreversible process, making the system less heterogeneous. Understanding the effect of these mechanisms on the final network topology would further enrich the range of possibilities to generate different microscopic architectures.

The ultimate aim of an effective multi-scale modeling approach is to transfer the knowledge obtained at a smaller scale to the next level of description. In this respect, a very promising step forward would be to use the results obtained through accurate monomer-resolved models as a guidance to develop more coarse-grained models that include some internal degrees of freedom and thus naturally include many-body interactions. A few examples of this kind have been recently proposed and could be potentially promising for microgels. Among these, we recall the Voronoi model widely used to describe biological tissues, which encodes the particle elasticity in the Hamiltonian~\cite{li2018role} and the liquid drop model, which has been used to calculate the phase diagram of polymeric particles at large compressions~\cite{doukas2018structure}. Finally, explicit models that are able to capture particle deformation and shrinking will be crucial to tackle by simulations the problem of the glass transition at high enough particle volume fractions. A recent study put forward a first model in this direction~\cite{gnan2018microscopic}.

As a final note, we stress that, although the present review has focused on non-ionic microgels, many of the issues that we have discussed will be  also relevant to accurately model ionic microgels. However, for highly charged systems, the electrostatic force usually dominates over the other contributions, and thus the microscopic details become somewhat less important for the swelling transition~\cite{fernandez2000charge}. In this context, important questions linked to the counter-ion distribution~\cite{kobayashi2017polymer,schneider_ions} and to the chemical equilibria of ions in weak polyelectrolyte nanogels~\cite{sean2017computer} have been the subject of recent work. Interestingly, since PNIPAM microgels are also  weakly charged~\cite{pelton1989particle, daly2000temperature, utashiro2017zeta}, with the effect of such a charge showing up close to the VPTT\cite{truzzolillo2018overcharging}, some of these results may be broadly relevant. Understanding the interplay between the electrostatic interactions and the onset of the swelling transition will be crucial to develop coarse-grained models that can be used across (and beyond) the VPT.

\section{Conclusions}
\label{sec:conclusions}

Here we have presented an overview of the numerical methods and models that have been used to investigate the behaviour of non-ionic microgels at many different length-scales, from the molecular level up to much more coarse-grained descriptions. We have put particular emphasis on the modelling of PNIPAM (thermoresponsive) microgels, which are increasingly being used as model systems to investigate fundamental problems in condensed-matter physics, but many of the results and methods reported here can be extended to other types of microgels. We have highlighted the inherent multi-scale nature of microgels. Indeed, part of the behaviour of individual microgels, such as the temperature at which thermoresponsive microgels deswell, can be directly traced back to the properties of the polymeric repeating units they are made of. However, other fundamental quantities, such as the swelling ratio or the single-particle elastic moduli, depend on the mesoscopic architecture of the network. The bulk (macroscopic) behaviour, in turn, is controlled by all these properties as well as by external parameters such as temperature, pH and salt concentration. Modelling microgels is thus a multi-faceted challenge that cannot be addressed with a single tool or technique. 

We have shown that in the last years there has been a flourishing of numerical studies on microgels. However, most of the effort has been devoted towards building realistic microgels to understand the single-particle rather than the bulk properties. In the meantime, the fast development of synthesis, imaging and scattering techniques have made it possible to experimentally probe dense suspensions of microgels to shed light on important open issues such as the jamming and glass transitions. Time is ripe now for the numerical community to catch up and use the knowledge gained by investigating the single-particle properties as a guidance for developing models which are simple enough to allow for bulk simulations but still take into account some of the details of the inner structure of microgels, as prefigured in the Perspectives section. A first challenging task for such a model would be to provide a microscopic explanation of the significant effect of the softness on the dynamical behaviour of suspensions of microgels~\cite{mattsson2009soft,nigro2018structural}.

\section*{Acknowledgements}
We acknowledge support from the European Research Council (ERC Consolidator Grant 681597, MIMIC). We thank our collaborators M. Bergman, F. Camerin, E. Chiessi, J. Crassous, F. Lo Verso, G. Del Monte, A. Ninarello, P. Schurtenberger and F. Sciortino with whom some of the work reviewed here has been carried out and for many valuable discussions.

\bibliography{library}

\end{document}